\documentclass[letter]{aa} 

\usepackage{graphicx}
\usepackage{txfonts}
\usepackage{hyperref}
\newcommand{\pl}{$\pm$}

\begin{document} 


   \title{Lifting the curtain: The Seyfert galaxy Mrk 335 emerges from deep low-state in a sequence of rapid flare events}
  \titlerunning{Mrk 335 in outburst after deep multi-year low state}


   \author{S. Komossa\inst{1}
             \and
         D. Grupe\inst{2}
                   \and
         L.C. Gallo\inst{3}
                   \and
         P. Poulos\inst{2}
          \and 
          D. Blue\inst{3}
          \and
         E. Kara\inst{4}
          \and
          G. Kriss\inst{5}
          \and
          A.L. Longinotti\inst{6}
          \and
          M.L. Parker\inst{7} 
          \and
          D. Wilkins\inst{8}}
         
   \institute{Max-Planck-Institut f\"ur Radioastronomie, Auf dem H{\"u}gel 69, 53121 Bonn, Germany\\\
              \email{astrokomossa@gmx.de}
         \and Dept. of Physics, Earth Science, and Space System Engineering, Morehead State University, 235 Martindale Dr, Morehead, KY 40351, USA
          \and  Department of Astronomy and Physics, Saint Mary’s University, 923 Robie Street, Halifax, NS, B3H 3C3, Canada    
          \and  MIT Kavli Institute for Astrophysics and Space Research, Cambridge, MA 02139, USA 
          \and Space Telescope Science Institute, 3700 S. Martin Drive, Baltimore, MD 21218, USA 
          \and Instituto Nacional de Astrofísica, Óptica y Electrónica INAOE- CONACyT, México 
          \and European Space Agency (ESA), European Space Astronomy Centre (ESAC), E-28691 Villanueva de la Canada, Madrid, Spain 
          \and Kavli Institute for Particle Astrophysics and Cosmology, Stanford University, 452 Lomita Mall, Stanford, CA 94305, USA 
             }

   \date{Received 3 August 2020; accepted 28 September 2020; journal reference: A\&A 643, L7}

 
  \abstract
   {The narrow-line Seyfert 1 galaxy Mrk 335 was one of the X-ray brightest active galactic nuclei, 
   but it has systematically faded since 2007.}
   {We report the discovery with Swift of a sequence of bright and rapid X-ray flare events
   that reveal the emergence of Mrk 335
   from its ultra-deep multiyear low state.  }
   {Results are based on our dedicated multiyear monitoring of Mrk 335 with Swift.}
{ Unlike other bright active galactic nuclei, the optical--UV is generally not correlated with the X-rays in Mrk 335 
on a timescale of days to months. 
This fact either implies the absence of a direct link between the two emission 
components; or else it implies that the observed X-rays are significantly affected by (dust-free) absorption 
along our line of sight. 
The UV and optical, however, are closely correlated at the 99.99\% confidence level. The UV is 
leading the optical by $\Delta t= 1.5\pm{1.5}$d.
The Swift X-ray spectrum shows strong deviations from a single power law in all brightness 
states of the outbursts, indicating that significant absorption or reprocessing is taking place. 
Mrk 335 displays a  softer-when-brighter variability pattern at intermediate X-ray count rates, 
which has been seen in our Swift data since 2007 (based on a total of 590 observations). This pattern breaks down 
at the highest and lowest count rates.}
{We interpret the 2020 brightening of Mrk 335 as a decrease in column density and covering factor
of a partial-covering absorber along our line of sight in the form of a clumpy accretion-disk wind 
that reveals an increasing portion of the intrinsic emission of Mrk 335 from the disk and/or corona 
region, while the optical emission-line regions receive a less variable spectral energy 
distribution. 
This then also explains why Mrk 335 was never seen to change 
its optical Seyfert type (not `changing look') despite its 
factor $\sim$50 X-ray variability with Swift.}


   \maketitle
%

\section{Introduction}

Active galactic nuclei (AGN) in extreme minima or maxima states provide us with 
important insight into the physics of the black hole and accretion disk region. 
The group of narrow-line Seyfert 1 galaxies (NLS1 hereafter) is known to exhibit 
particularly strong flux and spectral variability in X-rays (reviews 
by \citet{Gallo2018a} and \citet{Komossa2018}, and references therein).   
Mrk 335 \citep{Markarian1977} is a member of the class of NLS1 
galaxies at redshift $z=0.0258$ with a black hole mass of $M_{\rm BH} = 2.7 \times 10^7$ M$_{\odot}$ 
\citep{Grier2012}. Its optical spectrum is characterized by strong high-ionization coronal 
emission lines with ionization states of up to [FeX] \citep{Grupe2008} that 
imply the presence of a strong EUV-X-ray continuum. 
Mrk 335 was detected as a bright and variable 
UV emitter in early International Ultraviolet Explorer (IUE) and Hubble 
Space Telescope (HST) observations \citep[e.g.,][]{Dunn2006, Longinotti2013}.

For a time, this was one of the X-ray brightest AGN 
\citep{Tananbaum1978, Pounds1987, Grupe2001, Bianchi2001}. Then the X-ray flux of Mrk 335 dropped dramatically
in 2007 \citep{Grupe2007, Grupe2008}. 
Multiple epochs of exceptional flaring and dipping on its way to and beyond this 
low state motivated several deeper spectroscopic observations with XMM-Newton, Suzaku, NuSTAR, and the HST 
\citep{Grupe2012, Gallo2013, Kara2013, Longinotti2013, Parker2014, Gallo2015, 
Wilkins2015, WilkinsGallo2015, Komossa2017, Gallo2018, Gallo2019, 
Longinotti2019, Choudhury2019, Parker2019}.
These observations have shown that Mrk 335 is well fit with varying contributions of blurred reflection
and (ionized) absorption.
The reflection model explains diminished X-ray continuum emission with changes in a corona 
that has collapsed in toward the black hole, and sometimes forms a collimated outflow in
X-ray flare states. 
The absorption model explains several epochs of flares and fades with changes in ionization 
state and covering factor of the absorbers, which were detected with the XMM-Newton 
reflection grating spectrometer (RGS) and in the UV with the HST. 

We report the emergence of Mrk 335 from its recent ultra-deep low state 
(which lasted from late 2017 to at least January 2020), based on 45
new observations with the Neil Gehrels Swift observatory 
\citep{Gehrels2004} each in all seven wavebands from optical to X-rays. These 
are part of our long-term monitoring of Mrk 335, which started in May 2007. When Mrk 335 was observed on 2020 May 16, it was found already at a flux level that exceeded all observations since February 2018. 
In particular, it was very bright in the UV as well and appeared at a similar UV flux 
level as seen last during the flare at the end of 2011. Because of this outburst in 
May-June 2020, we started an intensive monitoring campaign with Swift, and we report the results here.
We use a cosmology \citep{Wright2006} with 
$H_{\rm 0}$=70 km\,s$^{-1}$\,Mpc$^{-1}$, $\Omega_{\rm M}$=0.3 and $\Omega_{\rm \Lambda}$=0.7 throughout this paper.

\section{Data analysis}

\subsection{Swift XRT} 
\begin{figure}
\includegraphics[clip, trim=1.3cm 5.6cm 1.3cm 2.6cm, width=\columnwidth]{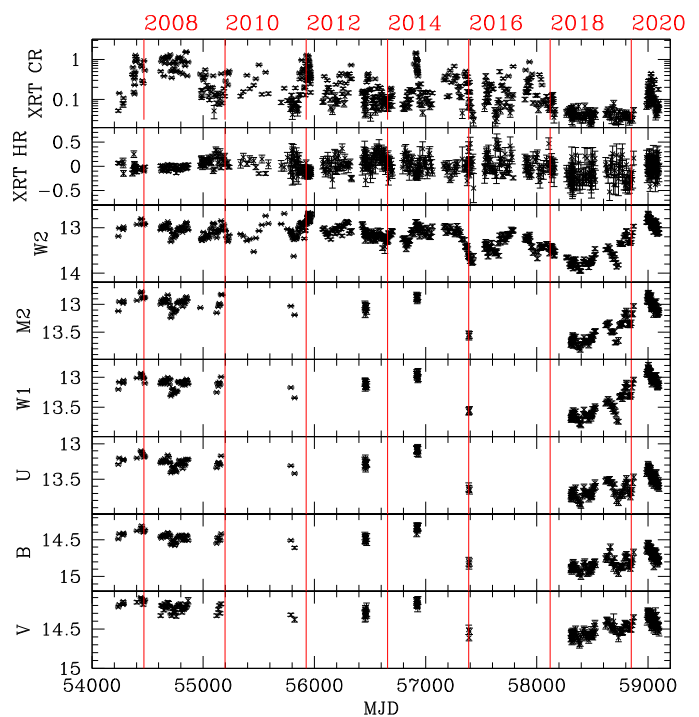}
    \caption{Long-term Swift XRT and UVOT light curves of Mrk 335 from 2007 May
    to 2020 July.  
    From top to bottom: XRT count rate, hardness ratio, and UVOT Vega magnitudes corrected for 
    Galactic extinction. The vertical red lines mark the beginning of each second year from 
    2008 to 2020.} 
\label{fig:lc-Swift}
\end{figure}

We have monitored Mrk 335 with Swift for more than 13 years since May 2007 with a cadence of typically 1-14 days, depending on the brightness and variability state of Mrk 335. Long gaps of several months occur each year when Mrk 335 is unobservable with Swift because of its proximity to the Sun. Mrk 335 came out of the Swift Sun constraint on 16 May 2020 \citep{Grupe2020} and we resumed our monitoring (Fig. \ref{fig:lc-Swift} and Fig. \ref{fig:lc-zoom}).  
The Swift X-ray telescope \citep{Burrows2005} was operating in photon-counting mode \citep{hill2004} 
with typical exposure times of 1-2 ks (Tab. \ref{tab:obs-log}).

X-ray count rates were determined using the online XRT product tool at the Swift data centre in Leicester \citep{Evans2007}{\footnote{ \url{https://www.swift.ac.uk/user_objects/}}}. 
For a dedicated spectral analysis of select epochs, source photons were extracted within a circle of radius 59\arcsec , and 
the background was determined in a nearby circle with a radius of 236\arcsec.  
During the faintest states, spectral fitting of individual spectra is not possible, and we instead measured hardness ratios, defined as $HR = \frac{hard-soft}{hard+soft}$, where {\it soft} and {\it hard} are the counts in the 0.3-1.0 and 1.0-10 keV energy bands, respectively. $HR$ was determined by applying the Bayesian estimation of hardness ratios, BEHR \citep{park06}{\footnote{ \url{http://hea-www.harvard.edu/AstroStat/BEHR/}}}. 

In order to obtain a high-state spectrum for spectral fitting of a better signal-to-noise ratio (S/N), we combined the 2020 June 23 -- July 1 Swift data when Mrk 335 was brightest (Target ID 13544, segments 011-016).
For comparison, we created a low-state spectrum when Mrk 335 was in its lowest state with data from May 2019 (Target ID 33420, segments 218-221).
These are the two primary data sets used for spectral fitting.
In both cases we created new ancillary response files (arfs) by adding the arfs of the single spectra weighted by their exposure time using the {\sc{ftool}} command {\it addarf}. The coadded X-ray spectra in the band (0.3-10) keV were then analyzed with the software package {\sc xspec} \citep[version 12.10.1f;][]{Arnaud1996}.

\subsection{Swift UVOT}
Mrk 335 was observed with the UV-optical telescope \citep[UVOT;][]{Roming2005}
either in W2 or in all six filters 
in order to obtain spectral energy distribution (SED) information of this rapidly varying AGN. 
Observations in each filter were coadded using the task {\em{uvotimsum}}.
Source counts in all filters were then extracted in a circular region with a radius 
of 5\arcsec ~, and the background was selected in a nearby region of radius 20\arcsec.
The background-corrected counts were then converted into VEGA magnitudes and 
fluxes based on the latest calibration as described in \citet{Poole2008} and \citet{Breeveld2010}. 
The task {\em{ uvotsource}} was used to measure the magnitudes and flux densities.  
A Galactic reddening correction was applied to the 
UVOT data, with a value of $E_{\rm{(B-V)}}$=0.035 \citep{Schlegel1998}, with a correction factor in each filter 
according to Eq. (2) of \citet{Roming2009} and making use of the reddening curves of \citet{Cardelli1989}. 

\section{Results}

\subsection{Light curves and DCF analysis}
Fig.\ref{fig:lc-Swift} displays the long-term light curve of Mrk 335 in X-rays as 
well as in the optical/UV since the start of our Swift monitoring program in May 
2007, when it was discovered in an unusually low X-ray flux state \citep{Grupe2007}. 
 Figure \ref{fig:lc-zoom} highlights the 2020 evolution in all bands. 

Mrk 335 exhibited several short-duration X-ray flares over the last 13 years 
when it was in a low state overall. The new rise in X-ray flux 
after the long period of very low activity of more than two years is remarkable; it is 
unprecedented in Mrk 335. The systematic rise 
in the UV flux since about mid-2018 is even more remarkable. 

Fig. {\ref{fig:lc-zoom} shows the Swift results since we resumed our monitoring program on 2020 May 16. 
This light curve shows multiple rapid flare events in X-rays, including one rapid 
rise by a factor of $\sim$5 within a week, and another rise by a 
factor of 2 within a day. Changes in the X-ray spectral shape as shown 
in the hardness ratio light curve are apparent. The XRT count 
rate before May 2020 was at a very low level around 0.05 cts s$^{-1}$ for more than 
two years (and it is still variable, therefore not host dominated),
and now (at the peak of the flare in late June 2020) it reached a value that is a factor of 10 higher. 
The brightest state in the UV (W2) was detected on 2020 May 31 with a Galactic-reddening 
corrected magnitude of 12.68\pl0.04. This is the highest state since the UV flare seen in December 2011. 

In order to search for correlations and time delays between different bands during the 2020 outbursts, 
a discrete correlation function \citep[DCF;][]{Edelson1988} analysis of the UVW2 and B and X-ray 
light curves was carried out following \citet{Gallo2018} with Monte Carlo confidence intervals. 
The UV band is closely correlated with the optical band, such that the optical is following 
and the UV leading 
by $\Delta t \simeq 1.5$d with a confidence interval of 3 to 0 days at the 99.99\% level. 
However, no correlation between the UV and X-rays is found. 

\begin{figure}
\includegraphics[trim=1.5cm 5.6cm 1.3cm 2.6cm, width=9.1cm]{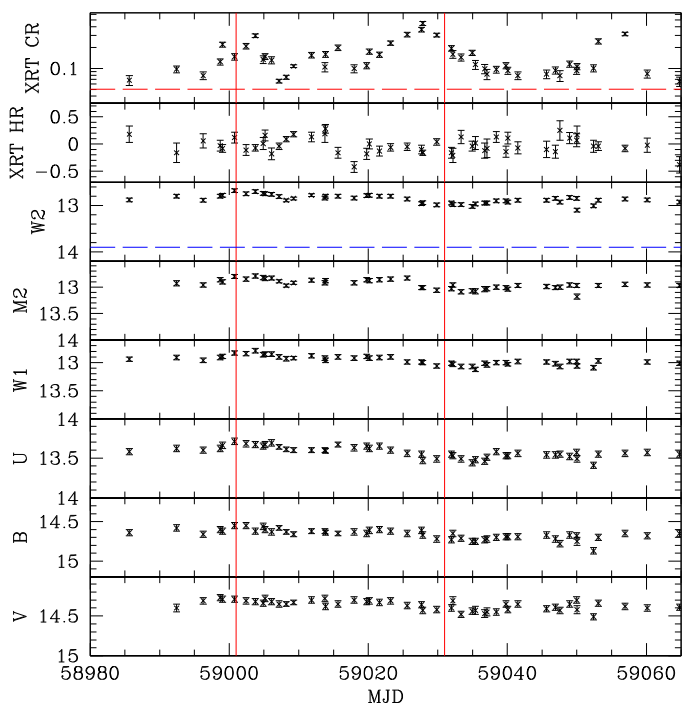}
    \caption{Swift outburst light curve of Mrk 335 since 2020 May 16 (units as in Fig. \ref{fig:lc-Swift}). 
The horizontal dashed red line marks an X-ray count rate of 0.05 cts/s (the lowest recorded during the 2018-2019 
low state was 0.03 cts/s), while the horizontal dashed blue line marks the lowest value of the W2 magnitude 
measured with Swift (in 2018). The vertical red lines mark 2020 June 1 and July 1. 
    } 
    \label{fig:lc-zoom}
\end{figure}

\subsection{X-ray spectral analysis}

Spectra were first fit with single power laws of photon index $\Gamma_{\rm X}$ 
adding Galactic absorption with a column density $N_{\rm H, Gal}=3.96\times10^{20}$ cm$^{-2}$ \citep{kalberla05}. 
Because this model did not provide good spectral fits and left strong systematic residuals in all fits (Fig. \ref{fig:xray_spec}), we then applied either an ionized or neutral partial covering absorber (Tab. \,\ref{tab:spec-fits}).
The single-component partial covering model consistently requires a lower column density and covering fraction at  high state. Alternatively, the high-state spectrum is well fit by an ionized absorber. 
The 2019 low-state spectrum can also be well described by a simple powerlaw plus black 
body model, where the black body is taken to be representative of a soft excess, which
is dominated by a number of emission lines detected with the RGS in 
low state \citep{Longinotti2019, Parker2019} and not individually resolved with Swift. 
We have added such a component (with its parameters fixed to the low-state values) 
when fitting the high-state spectrum, but it does not provide a significant change in 
the overall fit results. 
Based on the simple partial covering absorber fit to the 2020 high state, a (0.3-10) keV X-ray 
luminosity of $8.2 \times 10^{43}$ erg/s is obtained, or an X-ray Eddington ratio 
of $L_{\rm X}/L_{\rm Edd} \approx 0.02$, while for the 2019 low state, the intrinsic X-ray 
luminosity is a factor of 1.6 lower when the same model description is adopted.

Spectral fits were also performed on a 2020 low state and intermediate state (Fig. 3). However, as
Mrk 335 varies strongly during the 2020 time interval (unlike during the deep May 2019 low state),
spectral fits are less reliable and still come with large errors, and at any given time 
interval a mix of cold and ionized absorbers may contribute. 
These cannot be disentangled with Swift, however. 
Applying the simple partial covering model
to the 2020 low state,
we find  $N_{\rm H} = 11.32^{+4.62}_{-2.97}$ cm$^{-2}$ 
, while for the intermediate state, we find $N_{\rm H} = 4.32^{+1.22}_{-0.88}$ cm$^{-2}$. 

\subsection{Hardness ratio variability} 
We inspected the long-term (since 2007) and short-term (2020) $HR$ variability of 
Mrk 335 (Fig. \ref{fig:HR}) including all data points until 2020 July 19. 
Three regimes are apparent: a softer-when-fainter pattern at lowest count rate $CR$, 
a softer-when-brigther trend at intermediate $CR$, and an almost constant $HR$ at highest $CR$.
These three regimes have also been identified in other absorbed AGN that are intrinsically bright \citep[e.g.,][]{Connolly2014}.   
Results of a regression analysis (available in {\sc{midas}}) of the three variability regimes are shown in Fig. \ref{fig:HR}. A correlation analysis at $CR <0.1$ cts/s gives $r_s = +0.5, T_S = 8.7$,  
and $P<10^{-8}$ ($N$=234). At $0.1 < CR < 0.25$ cts/s, 
$r_s = -0.27, T_S=-3.7$, and $P=0.00014$  ($N$=177), while at $CR>0.25$ cts/s, $r_s = -0.205, T_S = -2.7$, and $P = 0.0038$ ($N$=174), where $r_s$ is the Spearman rank order correlation coefficient, 
$T_S$ is the Student T-test, $P$ is the probability that the distribution is just random, and $N$ is the number of data points.

\begin{table}
\scriptsize
        \centering
        \caption{
        X-ray spectral fit results of Mrk 335 (see Sect. 3.2).
        Columns are as follows: (1) Models: pl = powerlaw, bbdy = black body, zpcf = partial covering absorber, and zxipcf = ionized absorber model.
        (2) Absorption in units of 10$^{22}$ cm$^{-2}$; (3) power-law photon index; (4) $kT$ in units of keV of the black body, or ionization parameter log $\xi$ of the ionized absorber;
        (5) covering factor; and 
        (6) goodness of fit, reduced $\chi^2$. If no errors are reported, the quantity was fixed. For all fits the column density at $z=0$ was fixed to the Galactic value of 3.96$\times 10^{20}$ cm$^{-2}$.}
        \label{tab:spec-fits}
        \begin{tabular}{lccccc}
        \noalign{\smallskip}
                \hline
                \noalign{\smallskip}
                model & $N_{\rm H}$ & $\Gamma_{\rm X}$ & $kT$ or log $\xi$ & 
                $\eta$ &
                $\chi{^2}_{\rm red}$ \\
                (1) & (2) & (3) & (4) &  (5) & (6)   \\
                \noalign{\smallskip}
                \hline
                \noalign{\smallskip}
                \multicolumn{6}{c}{high state, 2020 June 23 -- July 1}  \\
                \noalign{\smallskip}
                \hline
                \noalign{\smallskip}
                pl & & 2.49\pl0.06 & &  & 3.2 \\
                zpcf * pl & 6.3$^{+3.3}_{-1.9}$ & 2.95$^{+0.14}_{-0.14}$ & &  0.86$^{+0.04}_{-0.05}$ &  1.4 \\
                bbdy+(zpcf * pl)$^1$ & 4.3$^{+1.7}_{-1.1}$ & 2.97\pl0.20 & 0.15 & 0.88$^{+0.04}_{-0.05}$ &  1.5\\
                zxipcf * pl & 16.1$^{+4.2}_{-4.9}$ & 2.3$^{+0.1}_{-0.1}$ & 2.1$^{+0.2}_{-0.1}$ & 0.87$^{+0.07}_{-0.06}$ & 1.1 \\
        \noalign{\smallskip}
                \hline
        \noalign{\smallskip}
                \multicolumn{6}{c}{deep low state, 2019 May$^2$}     \\
                \noalign{\smallskip}
                \hline
                \noalign{\smallskip}
                pl & & 2.34\pl0.20 & & & 1.6 \\
                zpcf * pl & 13.95$^{+8.62}_{-5.06}$ & 3.07\pl0.21 &  & 0.94$^{+0.03}_{-0.06}$ & 1.0  \\
                pl + bbdy & & 0.50$^{+0.51}_{-0.55}$ & 0.15\pl0.02 &  & 0.9  \\
                \noalign{\smallskip}
                \hline
        \end{tabular}

$^1$ bbdy parameters fixed at low-state value

$^2$ Fits using Cash statistics \citep{cash79}  
\end{table}

\begin{figure}
\includegraphics[clip, trim=1.0cm 1.0cm 3.0cm 2.6cm, width=7.1cm]
{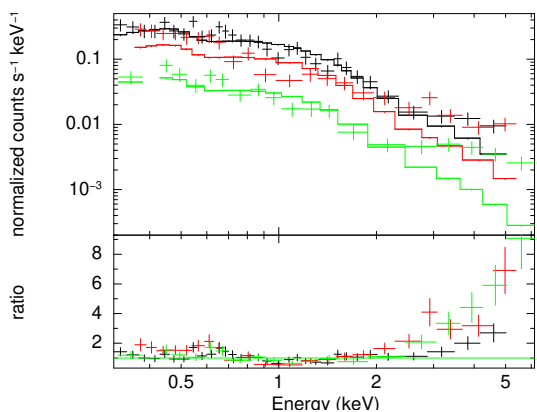}
    \caption{
    Selected Swift spectra of Mrk 335 in 2020 May--July. 
    The 2020 June high state is plotted in black, 
    an intermediate state from June 2 and 3 in red, and a low state from 2020 
    (data from May 16, 23, and 27, and June 7) in green. All spectra are fit with 
    a single power-law model with Galactic absorption, which displays strong 
    fit residuals especially at low and high energies.}
    \label{fig:xray_spec}
\end{figure}

\subsection{Spectral energy distribution} 
The SEDs during the high state in May-June 2020 in comparison with a low -state data set in May 2019 are shown in Fig. \ref{fig:SED}. The results 
show that the optical-UV spectrum has become bluer. While the W2-U color 
during the low state in 2019 was --0.03\pl0.06, it was --0.61\pl0.06 on 2020 May 31. 
In X-rays,  Mrk 335 has not only become brighter, but the spectrum has 
changed as well (see Sect. 3.2).

\section{Discussion}

The long-term Swift light curve of Mrk 335 reveals that its deepest 
minimum-state in the UV was reached in 2018, when the X-ray emission was 
also at a deep and long-lasting low state, with little variability left 
in X-rays (Fig. \ref{fig:lc-Swift}). 
Except for one deviating epoch in 2019, and with some short-time variability 
superposed, the UV has been on a steady rise since its 2018 minimum, 
reaching and exceeding its former Swift} peak value (2011) again in 2020 May-June 
(Fig. \ref{fig:lc-Swift}){\footnote{at no time during the Swift monitoring 
did the UV emission of Mrk 335 reach the highest brightness level recorded 
with IUE in the 1980s (e.g., \citet{Dunn2006}, 
Fig. 1 of \citet{Grupe2008}, \citet{Longinotti2013}, Fig. 14 of \citet{Tripathi2020})}}.
Instead, the long-term X-ray light curve remained flat throughout 2018-2019, 
and we only detected a significant rise since 2020 May, with multiple peaks. 
Except that both reached a deep low-state in 2018 (\cite{Parker2019}, 
which was relatively short-lived in the UV, however; Fig. \ref{fig:lc-Swift}),
the variability observed with Swift in the UV and X-rays is not closely 
correlated on the timescale of days to months \citep[see also][]{Grupe2012, Gallo2018, Tripathi2020}:
multiple flares in the UV are not seen in X-rays and vice versa, as 
confirmed by our DCF analysis of the 2020 data. This behavior is markedly 
different from that of other bright type 1 
AGN  \citep{Shappee2014, Gardner2017, Buisson2017, McHardy2018, Edelson2019} and demands an explanation (see below).

\begin{figure}
\includegraphics[clip, trim=1.7cm 1.5cm 1.3cm 0.1cm, width=5.9cm, angle=270] {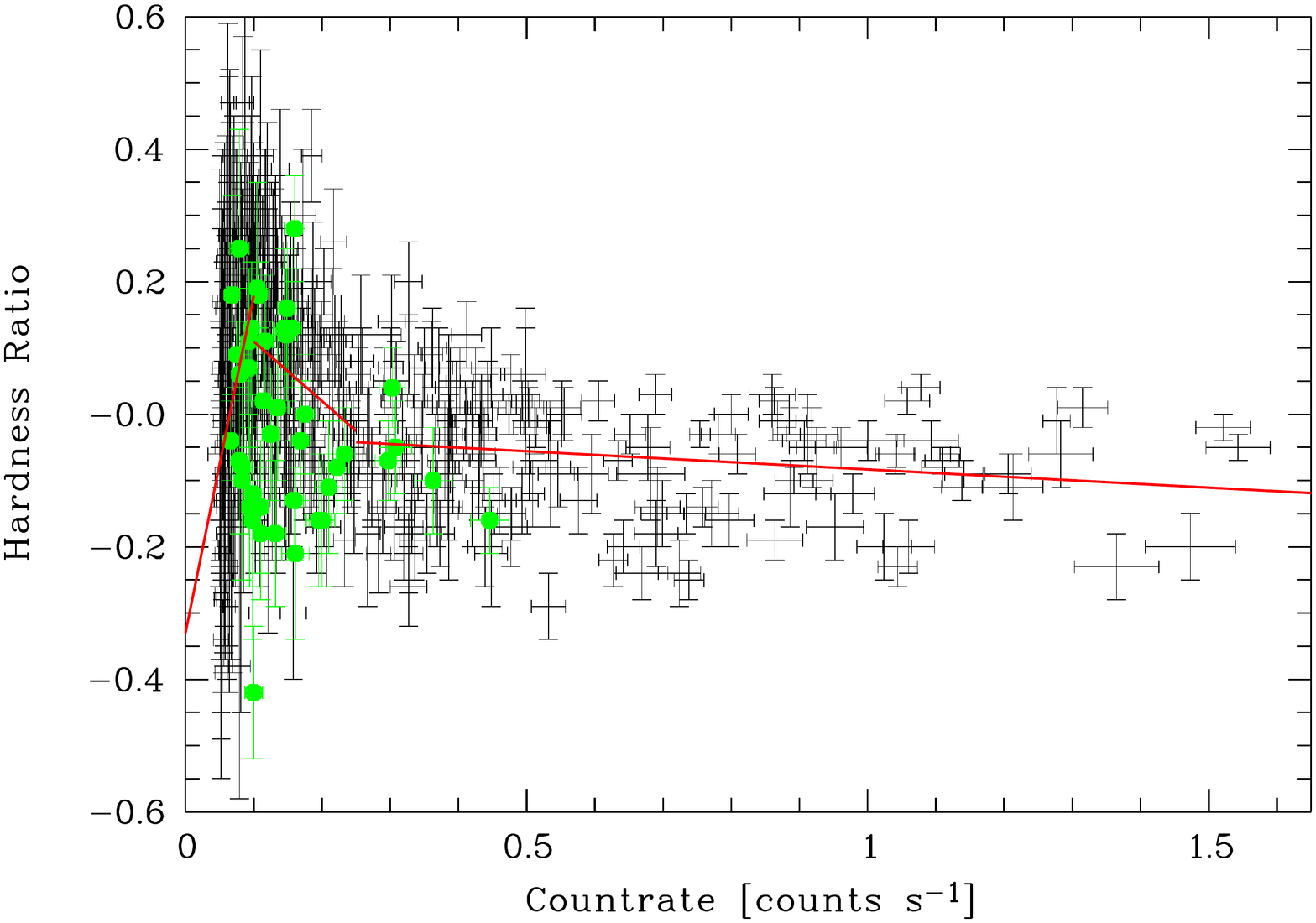}
\includegraphics[clip, trim=1.7cm 1.5cm 1.3cm 0.1cm, width=5.9cm, angle=270] {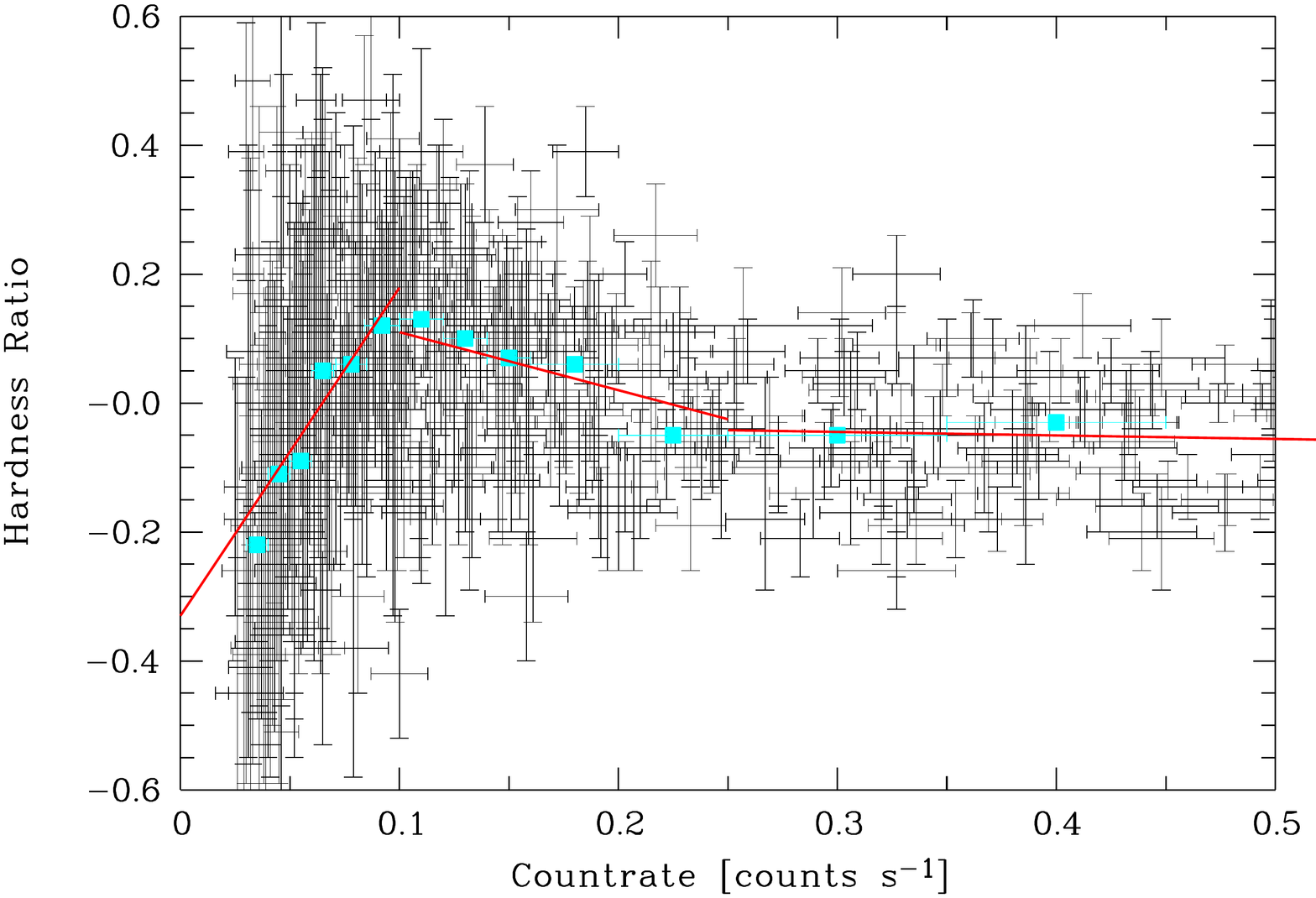}
    \caption{Hardness ratio vs. X-ray count rate for Swift XRT data of Mrk 335 since 
2007 (black). Three regimes are apparent: a softer-when-fainter trend at lowest count rate, a softer-when-brigther 
pattern at intermediate count rate, and a near-constant HR at highest count rate. The upper panel shows the full 
count-rate range. 
Green circles represent the 2020 outbursts. The lower panel zooms into the count-rate range up to 0.5 cts/s. The red 
lines represent the fits to the three regimes (Sect. 3.3), and the blue squares mark the median in each bin.}  
    \label{fig:HR}
\end{figure}
Several mechanisms are known to cause strong changes in the X-ray flux and spectral 
states of AGN: The first mechanism is changes in ionization state or geometry of ionized or 
neutral absorption that fully or partially covers the continuum source \citep[e.g.,][]{AbrassartCzerny2000, Turner2011, leighly2015}. 
Varying absorption is frequently seen in intermediate-type Seyfert galaxies whose variability timescales can be as short as hours or days, but can also occur in 
NLS1s \citep{Risaliti2011} when the absorber is part of a clumpy accretion-disk wind or 
resides within the BLR. 
The second mechanism is changes in the X-ray reflection of photons off the (inner) accretion disk 
\citep{Ross2005}. 
 This model has frequently been applied 
to explain short-time spectral changes in type 1 Seyfert galaxies by changes in the 
location and geometry of the corona. 
Third, on longer timescales, changes in the accretion rate including disk instabilities 
can cause high or low states in AGN
\citep[e.g.,][]{Ross2018, Czerny2019}. This last mechanism has been favored to explain 
some of the extreme changing-look AGN that also switch their Seyfert types. 

\begin{figure}
\includegraphics[clip, trim=1.9cm 1.5cm 2.0cm 0.8cm, width=6.0cm, angle=270] 
{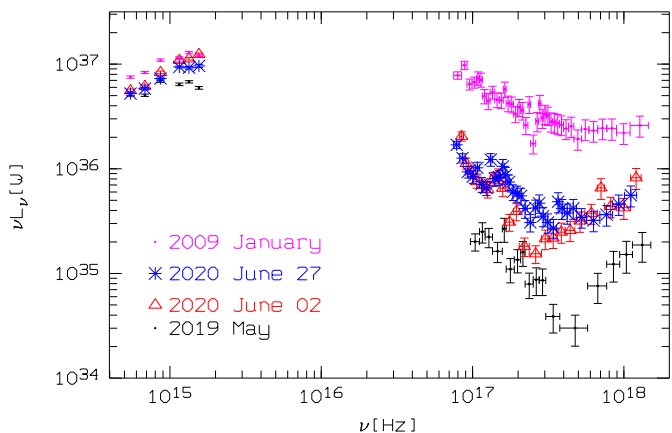}
    \caption{Observed Swift SED of Mrk 335 during 2020 June 2, June 27, and the deep low-state of 2019 May. 
    Note the X-ray flux deficit around (0.7-1.2)\,keV on June 2. For comparison, the Swift SED during the highest X-ray state from 2009 January is shown (purple).  
    }
    \label{fig:SED}
\end{figure}

In Mrk 335, the variability of the UV band is not correlated with the X-ray band on the 
timescale of days to months \citep[with the exception of an epoch in 2014;][]{Gallo2018}.
This rejects models that predict a close link between UV and X-rays: For instance, 
models where the observed X-rays are upscattered UV photons; or models where the UV 
is reprocessed emission from an X-ray corona heating the disk, which have successfully been applied 
to several well-observed AGN that show a close UV--X-ray 
correlation \citep{Shappee2014, McHardy2018}; or models where UV and X-rays 
are due to synchrotron radiation (unlikely for Mrk 335 even though it is a radio emitter).  
 
Furthermore , we can also exclude models of distant {\em{dusty}} absorbers that diminish the 
UV by extinction and simultaneously diminish the X-rays by absorption.  
However, {\em{dust-free}} absorption, arising in the accretion-disk region for instance 
in the form of a clumpy wind, is an efficient mechanism to cause X-ray spectral and flux 
changes while leaving the broad-band UV unaffected. And our spectral fits of a 
partial covering absorber consistently imply a decrease in covering factor 
and column density as the source brightens in X-rays. Our fit parameters of 
the ionization parameter, column density, and covering factor are consistent with predictions from models of clumpy disk winds \citep{Takeuchi2013}
that form via the Rayleigh-Taylor and radiation-hydrodynamic instability \citep{Shaviv2001} and 
which predict variability timescales on the  order of weeks. 

A partial covering scenario can also explain why Mrk 335 only shows a mild trend of softer-when-brighter (Fig. \ref{fig:HR}) because a partial coverer would preserve the spectral shape as it leaks different fractions of the same 
intrinsic continuum when the covering factor changes. The softer-when-fainter pattern at the very lowest 
count rates is likely due to a soft X-ray component in the form of multiple emission lines 
\citep{Longinotti2019, Parker2019} and a (distant) reflection component \citep{Gallo2019} which become increasingly apparent as the primary continuum diminishes. While the X-ray covering factor obtained from the current and 
previous (intermediate to low state) observations of Mrk 335 is high ($\sim$ 80-95\%), a partial covering scenario 
for Mrk 335 is independently favored by previous UV observations that revealed variable absorption lines, including 
Ly$\alpha$ at 20--30\%
covering fraction of the UV-emitting site \citep{Longinotti2019, Parker2019}. 

There is other evidence that the X-rays we observe along our line of sight are not representative of the 4$\pi$ emission: 
Optical reverberation mapping \citep{Grier2012} has shown that the Balmer line H$\beta$ and the 
high-ionization line HeII$\lambda$4686 of Mrk 335 follow the amplitude of the optical continuum 
variability{\footnote{the Balmer line emission is driven by the EUV beyond the Lyman limit, and the 
ionization potential of HeII is 54 eV}} and not the higher amplitude seen in X-rays. While $R_{\rm max}$ = 1.53 in optical 
continuum variability, and $R_{\rm max}$ = 1.55 in HeII (Grier et al. 2012; where $R_{\rm max}$ is the 
ratio of maximum to minimum flux in each light curve), we measured during the same time interval $R_{\rm max}$ = 5.44 in X-rays. 

Furthermore, no significant variability of the optical coronal lines in Mrk 335 was detected \citep{Grupe2008} despite 
strong changes in observed X-rays. As those lines are driven by the EUV-to-soft-X-ray part of the SED, the 
observations indicate that the emission lines saw a different (less variable) continuum. 
An absorption scenario can also explain the lack of intrinsic variability of the photoionized soft X-ray emission lines 
\citep{Parker2019}, and is consistent with the lack of IR variability of Mrk 335 \citep{Wright2010}. 

The question remains why Mrk 335 is no optical Seyfert-type changer{\footnote{as observed in an increasing number of other AGN \citep{MacLeod2016}}}
despite its high amplitude of X-ray variability (a factor $\sim$50 between highest and lowest state in our 
long-term Swift light curve). As described above, the emission-line regions seem to see a less variable EUV-X-ray 
SED. Furthermore, because the broad Balmer lines are intrinsically bright, even a drop of a factor of 2 still preserves the 
type 1 nature of Mrk 335. 

As Mrk 335 is a very bright AGN at high state, it may turn out to be a Rosetta stone in deciphering the various 
contributions of reflection, absorption, and intrinsic emission, which imprint their presence on AGN X-ray spectra.   
Following Mrk 335 closely as it emerges from low state will therefore provide us with an important opportunity of 
understanding the physical processes that drive long-term AGN state changes. 

\begin{acknowledgements}
We would like to thank Brad Cenko for approving our Swift observing requests and the Swift team for carrying out 
our observations, and our anonymous referee for very useful comments.  
This work made use of data supplied by the UK Swift Science Data Centre at the University of Leicester. ALL 
acknowledges support from CONACyT grant CB-286316. 
\end{acknowledgements}

%
%


\appendix
 \section{Log of observations}
\begin{table*}
       \centering
        \caption{Summary of Swift observations of Mrk 335 in 2020 May -- July. } 
        \label{tab:obs-log}
        \begin{tabular}{lcllcrrrrrrr}
                \hline
Target ID & Segment & obs start (UT) & obs end (UT) & MJD$^1$ & $t_{\rm XRT}^2$ & $t_{\rm V}^2$ & $t_{\rm B}^2$ & $t_{\rm U}^2$ &
$t_{\rm W1}^2$ & $t_{\rm M2}^2$ & $t_{\rm W2}^2$ \\
                \hline
33420 & 240 & 2020-05-16 15:15 & 2020-05-16 15:25 & 58985.63905 &  637 &  --- &  66 &  86 &  79 &   --- &  266 \\
      & 241 & 2020-05-23 09:34 & 2020-05-23 09:42 & 58992.43683 
&  480 &   38 &   38 &    38 &   78 &  105 &  154 \\ 
      & 242 & 2020-05-27 06:12 & 2020-05-27 06:29 & 58996.26405
&  995 &  79 &   79 &  79 &  157 &  249 &  315 \\
      & 243 & 2020-05-29 16:56 & 2020-05-29 17:10 & 58998.71010
&  834 &  70 &   70 &  70 &  141 &  171 &  280 \\  
      & 244 & 2020-05-30 00:55 & 2020-05-30 01:09 & 58999.04275
&  834 &  69 &   69 &  69 &  139 &  178 &  277 \\ 
      & 245 & 2020-05-31 18:19 & 2020-05-31 18:34 & 59000.76809 
&  897 &  73 &  73 &  73 & 145 &  205 &  292 \\ 
      & 246 &  2020-06-02 10:05 & 2020-06-02 10:15 & 59002.42337
&  587 &  45 &  45 &  45 &  90 &  149 &  181 \\
      & 247 & 2020-06-03 18:11 & 2020-06-03 18:28 & 59003.76332
&  999 &  81 &  81 &  81 & 163 &  226 &  328 \\ 
      & 248 & 2020-06-04 19:55 & 2020-06-04 23:11 & 59004.89769
&  974 &  77 &   77 &  77 &  155 &  215 &  311 \\ 
      & 249 & 2020-06-05 03:47 & 2020-06-05 04:01 & 59005.16214
&  844 &  69 &   69 &   69 &  137 & 192 &  273 \\
      & 250 & 2020-06-06 02:04 & 2020-06-06 02:20 & 59006.09137 
&  954 &  79 &  79 &  79 &  157 &  210 &  314 \\ 
      & 003 & 2020-06-11 20:23 & 2020-06-11 20:40 & 59011.85493
& 1009 &  82 &  82 &  82 & 165 &  229 & 330 \\
      & 005 & 2020-06-13 18:36 & 2020-06-13 18:54 & 59013.78176
&  932 &  77 &  77 &  77 & 154 &  207 &  307 \\     
      & 004 & 2020-06-13 20:23 & 2020-60-13 20:38 & 59013.85416
&  904 &  74 &  74 &  74 &  146 &  208 &  293 \\  
      & 006 & 2020-06-15 15:18 & 2020-06-15 15:31 & 59015.64189
&  739 &  56 &   81 &  81 & 164 & --- & 329 \\
      & 007 & 2020-06-17 23:05 & 2020-06-17 23:16 & 59017.96548
&  629 &  50 &  50 &  50 &  101 & 143 &  203 \\
      & 008 & 2020-06-19 18:12 & 2020-06-19 18:29 & 59019.76407
&  989 &  83 &  83 &  83 & 163 &  216 &  328 \\
      & 009 & 2020-06-20 03:44 & 2020-06-20 03:59 & 59020.16048
&  889 & 73 &  73 &  73 & 146 & 196 &  293 \\  
      & 010 & 2020-06-21 14:54 & 2020-06-21 15:11 & 59021.62666
&  974 & 81 &  81 &  81 & 160 & 213 & 322 \\
      & 011 & 2020-06-23 05:04 & 2020-06-23 05:21 & 59023.21670
& 1014 &  81 &  81 &  81 & 161 & 252 & 323 \\ 
      & 012 & 2020-06-25 14:18 & 2020-06-25 14:29 & 59025.59944
&  637 &  52 & 52  & 52 & 105 &  138 & 210 \\
      & 013 & 2020-06-27 15:39 & 2020-06-27 15:53 & 59027.65675
&  814 &  66 &  66 &  66 & 132 & 190 & 263 \\
     & 014 & 2020-06-27 20:22 & 2020-06-27 20:39 & 59027.85434
&  991 &  81 &  81 &  81 & 161 & 229 & 323 \\
     & 015 & 2020-06-29 20:11 & 2020-06-29 20:27 & 59029.84626 
&  946 &  77 &  77 &  77 & 153 & 219 & 307 \\     
     & 016 & 2020-07-01 23:09 & 2020-07-01 23:24 & 59031.96965
&  864 &  71 &  71 &  71 & 142 & 194 & 283 \\
     & 017 & 2020-07-02 04:03 & 2020-07-02 04:11 & 59032.17113
&  492 &  39 &  39 &  39 &  79 &  107 & 157 \\
     & 018 & 2020-07-03 08:37 & 2020-07-03 08:52 & 59033.36409
&  864 &  70 &  70 &  70 & 140 & 202 & 279 \\     
     & 019 & 2020-07-02 22:51 & 2020-07-04 23:06 & 59034.95704
&  887 &  72 &  72 &  72 & 143 & 206 & 287 \\
     & 020 & 2020-07-05 10:07 & 2020-07-05 10:15 & 59035.42399
&  477 &  37 &  37 &  37 &  76 &  110 & 151 \\
     & 021 & 2020-07-06 17:59 & 2020-07-06 18:07 & 59036.75174
&  482 &  37 &  37 & 37 &  75 & 116 & 148 \\
     & 022 & 2020-07-07 01:56 & 2020-07-07 02:05 & 59037.08344
&  524 &  40 &  40 &  40 &  80 & 135 & 159 \\     
     & 023 & 2020-07-08 09:57 & 2020-07-08 10:14 & 59038.42019
& 1011 &  81 &  81 &  81 & 160 & 253 & 320 \\
     & 024 & 2020-07-09 19:32 & 2020-07-09 19:48 & 59039.81892
&  944 &  79 &  79 &  79 & 157 & 224 & 315 \\
     & 025 & 2020-07-01 01:51 & 2020-07-10 02:07 & 59040.08225
&  964 &  78 &  78 &  78 & 156 & 229 & 312 \\     
     & 026 & 2020-07-11 12:38 & 2020-07-11 12:54 & 59041.53151
&  979 &  78 &  78 &  78 &  155 & 244 & 310 \\
     & 028 & 2020-07-05 15:34 & 2020-07-15 15:51 & 59045.65441
&  991 &  79 &  79 &  79 & 157 & 247 & 315 \\     
     & 027 & 2020-07-16 21:56 & 2020-07-16 22:09 & 59046.91742
&  894 &  72 &  72 &  72 & 144 & 212 & 287 \\
     & 029 & 2020-07-17 13:56 & 2020-07-17 14:04 & 59047.58317
&  452 &  36 &  36 &  36 &  74 &   92 & 147 \\
     & 030 & 2020-07-18 23:17 & 2020-07-18 23:32 & 59048.97506
&  889 &  73 &  73 &  73 & 146 & 198 & 293 \\
     & 031 & 2020-07-19 23:13 & 2020-07-19 23:28 & 59049.97230
&  884 &  72 &  72 &  72 & 144 & 206 & 287 \\     
     & 032 & 2020-07-20 00:46 & 2020-07-20 00:54 & 59050.03457
&  450 &  35 &  35 &  35 &  70 & 104 & 141 \\
     & 033 & 2020-07-22 10:01 & 2020-07-22 10:16 & 59052.42186
&  864 &  73 &  73 &  73 & 145 & --- & 290 \\
     & 034 & 2020-07-23 02:16 & 2020-07-23 02:26 & 59053.09802
&  649 &  51 &  51 &  51 & 102 & 153 & 205 \\     
     & 035 & 2020-07-26 22:37 & 2020-07-26 22:54 & 59056.94811
&  989 &  80 &  80 &  80 & --- & 239 & 317 \\    
     & 036 & 2020-07-30 03:10 & 2020-07-30 03:32 & 59060.13607
&  789 &  62 &  62 &  62 & 125 & 190 & 250 \\     
\hline
        \end{tabular}

$^1$ The Modified Julian Date is given for the middle of the observation.         

$^2$ All exposure times $t$ are given in seconds.
        
\end{table*}

\end{document}